\documentclass{article}
\usepackage{url}
\usepackage{graphicx}
\usepackage{amsmath}
\usepackage{amsfonts}

\begin{document}
\title{\large\bf Secure Communication Protocol for Smart Transportation Based on Vehicular Cloud
}
\author{Trupil Limbasiya\footnote{BITS, Pilani, Dept. of CS \& IS, Goa Campus, Goa, India, Email: ssahay@goa.bits-pilani.ac.in} \hspace{0.15mm} Debasis Das\footnote{IIT, Jodhpur, Rajasthan, India, Email: debasis@iitj.ac.in} and Sanjay K. Sahay\footnote{BITS, Pilani, Dept. of CS \& IS, Goa Campus, Goa, India, Email: ssahay@goa.bits-pilani.ac.in}}

\date{}

\maketitle

\begin{abstract}
	The pioneering concept of connected vehicles has transformed the way of thinking for researchers and entrepreneurs by collecting relevant data from nearby objects. However, this data is useful for a specific vehicle only. Moreover, vehicles get a high amount of data (e.g., traffic, safety, and multimedia infotainment) on the road. Thus, vehicles expect adequate storage device for this data, but it is infeasible to have a large memory in each vehicle. Hence, the vehicular cloud computing (VCC) framework came into the picture to provide a storage facility by connecting a road-side-unit (RSU) with the vehicular cloud (VC). In this, data should be saved in an encrypted form to preserve security, but there is a challenge to search for information over encrypted data. Next, we understand that many of vehicular communication schemes are inefficient for data transmissions due to its poor performance results and vulnerable to different fundamental security attacks. Accordingly, on-device performance is critical, but data damages and secure on-time connectivity are also significant challenges in a public environment. Therefore, we propose reliable data transmission protocols for cutting-edge architecture to search data from the storage, to resist against various security attacks, and provide better performance results. Thus, the proposed data transmission protocol is useful in diverse smart city applications (business, safety, and entertainment) for the benefits of society.
\vspace*{0.1cm}
~
\end{abstract}

\section{Introduction}
The vehicular ad-hoc network (VANET) is a specific environment for vehicle users to exchange messages (of traffic information, weather situations, road conditions, etc.) with other nearby vehicles and road-side-units (RSUs). The VANETs include two types of communications as V2I (vehicle-to-infrastructure) and V2V (vehicle-to-vehicle), which are done using dedicated short-range communications (DSRC) over the physical layer and the data link layer of the OSI model \cite{Kenney2011}. The VANET has various applications, e.g., road safety, driver assistance, payment, curve speed warning, lane change warning, emergency status, forward collision, store location, etc. Thus, an intelligent transport system (ITS) is an essential application of VANETs \cite{Karagiannis2011}, \cite{Al-Sultan2014}. There are two communication components, i.e., RSU and OBU (on-board-unit). An RSU is available on the road network to transmit information to nearby vehicles. An OBU is installed in a vehicle to transfer messages to nearby RSUs and other OBUs.  

By 2020, it is anticipated that 75\% car of the world will be enabled with web services. Hence, vehicle users can perform different day-to-day operations (entertainment, data sharing, payment, online shopping, social media, etc.) over the Internet in a vehicle. The Internet of Vehicles (IoV) structure was proposed to revolutionize existing research fields (wireless sensor, VANET, infrastructure, and mobile device) by connecting them with smart transportation using different communication technologies (i.e., DSRC, wireless access points (WAP), and 4G/5G). It is designed with five diverse communications, i.e., V2V, V2I, V2R (vehicle-to-RSU), V2M (vehicle-to-mobile device), and V2S (vehicle-to-wireless sensor). This architecture has different features for on the fly data transmissions, i.e., direct connection with the end-user, different communication types, extended communication range, advanced applications, and network/data awareness \cite{Gerla2014}, \cite{Kaiwartya2016}. All these IoV communications happen publicly, and therefore, security comes into the picture to preserve data and user security in terms of data verification, user authentication, message confidentiality, on-time data access, and non-repudiation. \cite{Sun2017}. Further, data is transmitted between two entities in the IoV architecture, and it is not saved anywhere for future usage. Consequentially, this data is useful for only two objects temporary. In other words, this data has limited usage for a specific time and two users only.

In the fast-growing world, the VCC architecture was proposed to fulfill storage requirement on-demand on the road. In the VCC structure, three components (vehicular cloud (VC), OBU, and RSU) are connected to enhance the vehicular communication system. Here, we have primarily three communications (vehicle-to-RSU, vehicle-to-vehicle, and RSU-to-VC) to transfer data over an insecure channel \cite{Lee2014}, \cite{Whaiduzzaman2014}, \cite{He2014}. Accordingly, vehicle users have transportation-related services using the \textit{VCC} architecture. After studying different VCC structures as discussed in \cite{Lee2014}, \cite{Whaiduzzaman2014}, \cite{He2014}, \cite{Mekki2017}, \cite{Fan2018}, and \cite{Jiang2018}, we understand that these architectures have limited scope in communication. Besides, the VCC framework has different security and performance challenges \cite{Yan2013}.

\begin{figure}[h]
  \centering
  \includegraphics[width=\linewidth]{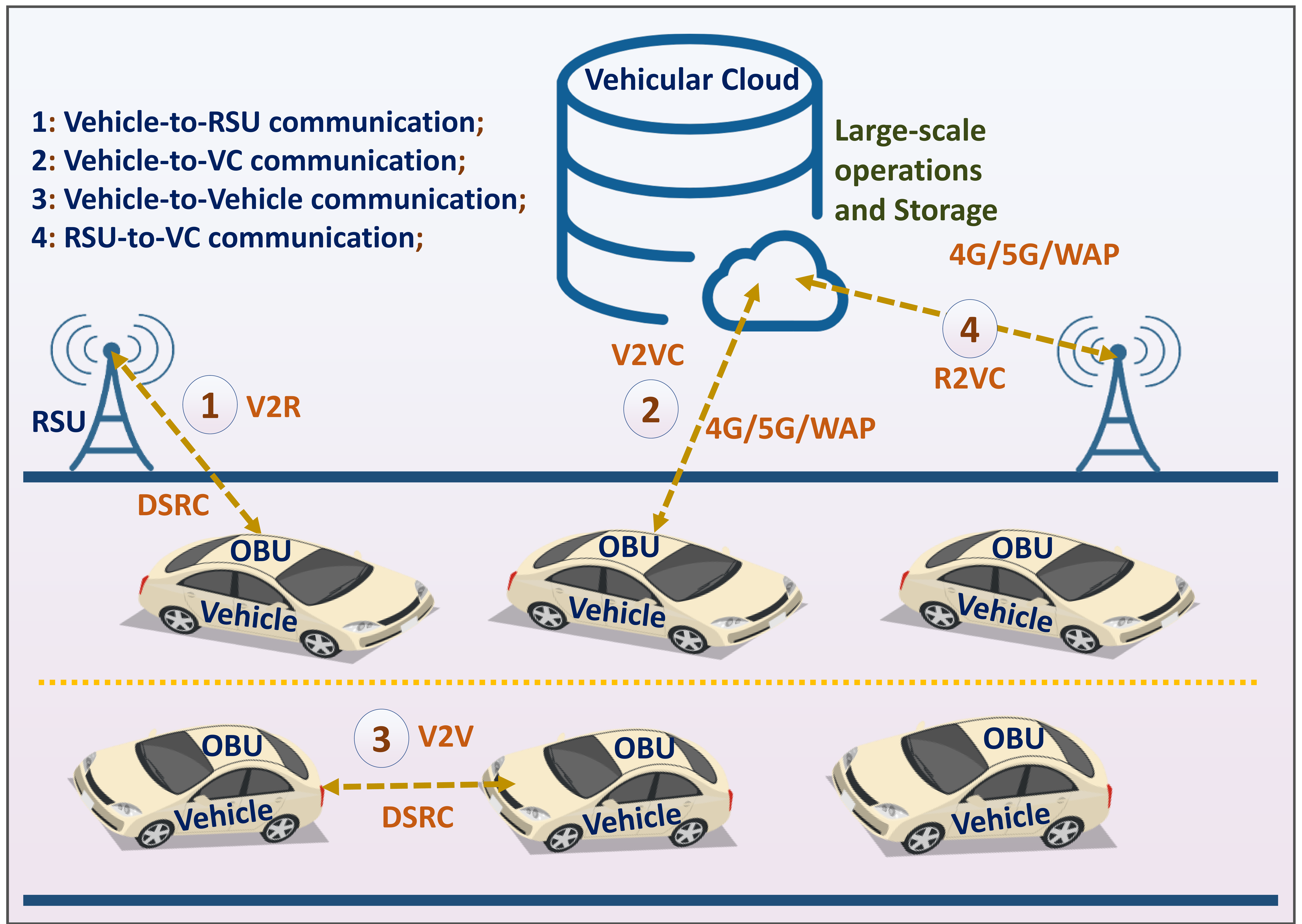}
  \caption{The proposed CVCC architecture.}
  \label{fig1}
\end{figure}

Therefore, vehicle users should have an extensive system to communicate with the VC systematically anytime around the world for a better society. Hence, we suggest with a comprehensive vehicular cloud computing (CVCC) system, which is responsible for exchanging data between different devices (OBU, RSU, VC, and government). Communications are carried out using DSRC, 4G/5G, and WAP technologies in the CVCC framework (see Figure \ref{fig1}). The VC is practiced to execute large-scale operations and to save meaningful data securely. Then, this data can be referred as an input for multiple purposes (e.g., road safety, emergency message transmission, optimization of traffic signals, toll plaza payment, future city development, tax payment, etc.). Therefore, the suggested VC architecture is a knowledge city for the society by connecting the automotive industry, ubiquitous technological systems, and smart governance system.

\section{Literature Survey}
In the last two decades, researchers have focused more on VANET due to different real-world applications. Further, there are many research and developments projects, which are already implemented in the EU, Japan, the United States, and other countries in the world but there are vital challenges (large-level data transmission and security) \cite{Al-Sultan2014}, \cite{Hartenstein2008}. Therefore, the communication scope of VANET is limited. Next, an adversary has an opportunity to perform various security attacks (e.g., impersonation, modification, replay, man-in-the-middle, plain-text, password guessing, session key disclosure, etc.) in VANETs.

To deal with multiple security problems, Zhang et al. \cite{Zhang2008} came up with a new communication system through an RSU, but this method failed to achieve security for attacks (man-in-the-middle and replay). Further, the protocol \cite{Zhang2008} expects a high amount of memory to save essential parameters, and the execution of this scheme is time-consuming. Similarly, researchers proposed different communication methods to deal with various problems in VANETs \cite{Manvi2017}. Accordingly, a good number of communication/authentication schemes are proposed for VANETs with distinct advantages and drawbacks. However, most of the systems are vulnerable to different attacks, and they need improvements in different performance measure.

Researchers (\hspace*{-2px} \cite{Lee2014}, \cite{Whaiduzzaman2014}, \cite{He2014}, \cite{Mekki2017}, \cite{Fan2018}, and \cite{Jiang2018}) suggested different VCC architectures to enhance the VANET system. Authors (\hspace*{-2px} \cite{Lee2014}, \cite{He2014}, and \cite{Jiang2018}) came up with a new VCC structure with different communications (three - V2V, V2R, and RSU2VC; two - V2R and RSU2VC; two - V2R and RSU2VC;) respectively. Next, other researchers (\hspace*{-2px} \cite{Whaiduzzaman2014}, \cite{Mekki2017}, and \cite{Fan2018}) suggested a VCC framework and did security analysis on them. Three communications (V2V, V2R, and RSU2VC) are available in \cite{Whaiduzzaman2014} and \cite{Mekki2017}. Four communications (V2V, V2R, V2M, and RSU2VC) are present in \cite{Fan2018}. Moreover, the performance measure (i.e., implementation time, communication overhead, storage cost, and energy consumption) of the VCC architecture is not discussed in \cite{Lee2014}, \cite{He2014}, \cite{Mekki2017}, and \cite{Jiang2018}. Next, other researchers (\hspace*{-2px} \cite{Whaiduzzaman2014}, \cite{Fan2018}, and \cite{Yan2013}) have discussed on the performance measure but they expect more resources to put into the practice. Further, Yan et al. \cite{Yan2013} only discussed on security by considering the VCC framework (of three communications - V2V, V2R, and RSU2VC).

According to \cite{Kaiwartya2016}, five communications are available in the IoV framework, and they are some data transmission schemes (\hspace*{-2px} \cite{Alam2015}, \cite{Liu2017}, \cite{Ruan2017}, \cite{Wu2017}) in which they have referred the IoV concept, but they have not designed data transmission protocols for all five communications. In \cite{Alam2015}, they proposed data transmission schemes for V2V, V2R, and vehicle-to-home. In \cite{Liu2017}, Liu et al. suggested a communication method for V2V only using different high-computational cryptographic techniques. As a result, this scheme \cite{Liu2017} is highly time-consuming, and thus, it is not appropriate in real-life applications. Ruan et al. \cite{Ruan2017} came up with an authentication protocol using wireless sensors, and data is transmitted to the server through a cluster head from a wireless sensor. Then, the server sends data to RSUs and OBUs. Hence, vehicle users cannot communicate directly with other IoV components. Further, this scheme \cite{Ruan2017} provides two communications, and it is insecure to multiple security attacks. In \cite{Wu2017}, they proposed a data transmission system for two communications (V2R and V2V). In V2V communication, a vehicle sends data to an RSU, and then, it transfers messages to nearby vehicles. Hence, the scheme \cite{Wu2017} does not provide a facility to exchange messages between vehicles directly. Further, this protocol does not resist to basic attacks (i.e., concatenation, modification, replay, and impersonation). After doing analysis on these schemes (\hspace*{-2px} \cite{Alam2015}, \cite{Liu2017}, \cite{Ruan2017}, \cite{Wu2017}), we understand that all these methods are not effective in security and performance.

\section{Problem Statement and Objectives}
From the literature survey, we understand that many of the communication schemes (of VANET, IoV, VCC) are inefficient for data transmissions due to its poor performance results. Further, these protocols cannot withstand against various security attacks, e.g., man-in-the-middle, replay, password guessing, session key disclosure, impersonation, chosen cipher-text, Sybil, modification, insider, etc. 

Next, data transmissions (in mobile computing) happen using cellular technology at 900/1800 MHz band with individual the capacity of 200 kHz, the data rate up to 2 Mbps and latency of 1.5 to 3.5 seconds. Further, a mobile device firstly sends data to a base station, and then, this base station transfers data to the receiver. Hence, there is no direct communication between (sender and receiver) in mobile computing. Thus, data transmissions are generally less efficient through mobile cellular technology in VANET. Moreover, wireless fidelity (Wi-Fi) is practiced for vehicular communications. In this, IEEE 802.11a provides a data rate of 54 Mbps at 5 GHz; IEEE 802.11b works at 2.4 GHz by delivering a data rate up to 11 Mbps, and IEEE 802.11g achieves 54 Mbps data rate at 2.4 GHz. However, Wi-Fi should be used in a limited manner because it has a communication range of 140 meters, which is not entirely suitable for the VANET structure. Vehicular communications are accurately carried out using DSRC with a range of 75 MHz of the spectrum (5.850-5.925 GHz) directly. The 75 MHz spectrum is divided into seven channels, and each channel has the capacity of 10 MHz. Further, DSRC supports a data rate of 27 Mbps, and latency is 200 microseconds. Hence, DSRC can be used to transmit messages efficiently on the road.

Further, data has limited usage for two entities temporary in VANETs and IoV. Next, the VCC provides a storage facility for OBU or RSU, but not both vehicular components (RSU and OBU) in the same architecture in the fast-growing world. It is essential to provide VC resources to both (OBU and RSU) because OBUs are installed in an individual vehicle and hence, vehicle users may have an opportunity to bogus data to the VC. Consequently, future operations might be misled using stored false data in the VC.

Generally, vehicular data transmissions happen over an insecure channel, and hence, it is also essential to achieve significant security level of the communication method. Thus, we need reliable communication protocols for the CVCC system. To deal with security and performance, we formulate our research and development objectives as follows.

\begin{itemize}
	\item Propose secure and cost-effective vehicular communication schemes (V2R, V2VC, V2V, and R2VC).
	
	\item Do security evaluations and performance analysis (execution time, communication and storage cost) on the proposed data transmission protocols.
	
\end{itemize}

\section{Research Methodology and Approach}
To overcome different performance and security drawbacks, we propose a reliable data transmission system for smart city applications by covering four different communications (V2V, V2R, RSU2VC, and V2VC). To design and develop these communications, we use cryptographic operations, i.e., one-way hash function ($h(\cdot)$), bit-wise XOR ($\oplus$), and concatenation ($||$), symmetric/asymmetric cryptography, bi-linear pairing, and elliptic curve cryptography (ECC). The VC data transmission system mainly consists of three phases as (1) initialization (2) registration and (3) message communication. Some existing data transmission protocols in VANETs are discussed in \cite{Zhang2008}, \cite{Liu2017}, \cite{He2015}, \cite{Wang2016}, \cite{Cui2018}.

\begin{enumerate}
	\item Initially, the registration authority (RA) generates and computes basic parameters to deploy RSUs on the road.
	
	\item In the registration process, different users enroll with the RA for future communications. Then, the RA puts an OBU and a tamper-proof-device (TPD) in a user's vehicle during the registration phase. An OBU consists of public parameters (of the RA), and a TPD includes a vehicle user's some secret computed parameters.

	\item The message communication phase includes three steps namely, (a) login and authentication at the sender side (b) message/request generation by the sender and (c) verification by the receiver and key-agreement at both sides (receiver and sender). 
	
	\begin{enumerate}
		\item A vehicle user inserts his/her identity and password in an OBU. Next, the system confirms the correctness of these credentials, and if valid, then it proceeds to the next step. Otherwise, it ends the session directly.
		
		\item The system generates a message request using a time-stamp and sends it to the receiver.
		
		\item The receiver (e.g., RSU/vehicle/VC) confirms the received message request and its sender. If valid then only, both (sender and receiver) generates a temporary session key using confidential credential(s), random nonce (agreed mutually), and time-stamp to start a communication with each other. 
	\end{enumerate}

\end{enumerate}

Our work plan is divided into four phases. The first phase is to design communication protocols for V2R, V2VC, V2V, and R2VC. In the second phase, the security analysis is done on these data transmission schemes using the random oracle model and security tools (AVISPA \cite{Vigano2006} and ProVerif \cite{Blanchet2018}). In the third phase, the proposed protocols are implemented on the test-bed set-up to check their execution time, communication overhead, storage cost, and energy consumption. An efficient data searching algorithm is designed to retrieve vital information from the VC (for a requested query), and this algorithm is verified for its efficiency to implement for vehicle users in the fourth phase.

We have proposed an efficient and secure communication scheme for V2R communication, as discussed in \cite{Limbasiya2019PC}, and this protocol work resists to various cyber-attacks, i.e., plain-text, man-in-the-middle, impersonation, modification, and replay. Further, the suggested protocol \cite{Limbasiya2019PC} is also feasible to verify a massive number of messages at a time precisely, and it performs excellently in the implementation cost, energy consumption, communication overhead, and storage cost compared to other relevant data transmission mechanisms.

According to the work plan, we have proposed a communication protocol \cite{Limbasiya2019WN} for vehicle users using the batch verification concept, in which a vehicle user can share vital information to nearby RSUs and other OBUs on the road. Further, this proposed scheme resists to impersonation, modification, replay, man-in-the-middle, password guessing, and stolen device attacks without using a TPD. Besides, the method \cite{Limbasiya2019WN} requires less computational resources, i.e., execution cost, communication overhead, storage cost, and energy consumption compared to relevant data transmission methods.

We came up with an effective V2V communication scheme \cite{Limbasiya2019ISJ} using a one-way hash function, in which a vehicle user sends meaningful information to nearby OBUs over a common channel. Moreover, this scheme is designed to preserve security requirements (i.e., authentication and integrity) in the communication system, and we have discussed the security proof of the proposed scheme. Thus, it is resistant to different security attacks, e.g., modification, replay, concatenation, impersonation, password guessing, man-in-the-middle, and stolen OBU attacks. Besides, it can be implemented without using a TPD with less computational resources for V2V communications. Hence, the protocol \cite{Limbasiya2019ISJ} is highly useful on the highway to transfer important messages to other vehicles.

We have proposed a V2R data transmission protocol \cite{Limbasiya2019PMC} using the EC concept and one-way hash function to exchange road-side and other relevant data between a vehicle user and an RSU. Moreover, it can withstand different security attacks, e.g., session key disclosure, replay, man-in-the-middle, impersonation, and modification. Further, it takes less computational resources for the implementation compared to other relevant communication protocols. Therefore, the scheme \cite{Limbasiya2019PMC} can be used for efficient and secure communications in smart transportation applications for a sustainable environment.

\end{document}